\documentclass{article}
\usepackage{spconf,amsmath,graphicx}
\usepackage{epstopdf}
\usepackage{array}
\usepackage{graphicx}
\usepackage{multirow}
\usepackage{color, soul}
\usepackage{cite}
\usepackage{url}
\usepackage{hyperref}
\usepackage{booktabs}
\usepackage{amsmath}
\usepackage{subfigure}
\usepackage{setspace}
\usepackage{makecell}
\usepackage{longtable}

\usepackage{amssymb}

\title{FluentEditor: text-based speech editing by considering \\ acoustic and prosody consistency
}
\name{Rui Liu$^{1}$\thanks{The research by Rui Liu is funded by the High-level Talents Introduction Project of Inner Mongolia University (No. 10000-22311201) and the Young Scientists Fund of the National Natural Science Foundation of China (NSFC) (No. 62206136). Haizhou Li is partly supported by National Natural Science Foundation of China (No. 62271432) and AME Programmatic Funding Scheme (Project No. A18A2b0046).}, Jiatian Xi$^{1}$, Ziyue Jiang$^{2}$, Haizhou Li $^{3,4}$}
\address{ $^1$ 
Inner Mongolia University, Hohhot, China $^2$ Zhejiang University, China
\\$^3$ Shenzhen Research Institute of Big Data, School of Data Science, \\The Chinese University of Hong Kong, Shenzhen (CUHK-Shenzhen), China \\ $^4$ National University of Singapore, Singapore\\
\small{liurui\_imu@163.com, x\_jiatian@163.com, ziyuejiang@zju.edu.cn, haizhouli@cuhk.edu.cn}
}
\begin{document}
%
\maketitle
\begin{abstract}
Text-based speech editing (TSE) techniques are designed to enable users to edit the output audio by modifying the input text transcript instead of the audio itself. Despite much progress in neural network-based TSE techniques, the current techniques have focused on reducing the difference between the generated speech segment and the reference target in the editing region, ignoring its local and global fluency in the context and original utterance.
To maintain the speech fluency, we propose a fluency speech editing model, termed \textit{FluentEditor}, by considering fluency-aware training criterion in the TSE training. Specifically, the \textit{acoustic consistency constraint} aims to smooth the transition between the edited region and its neighboring acoustic segments consistent with the ground truth, while the \textit{prosody consistency constraint} seeks to ensure that the prosody attributes within the edited regions remain consistent with the overall style of the original utterance.
The subjective and objective experimental results on VCTK demonstrate that our \textit{FluentEditor} outperforms all advanced baselines in terms of naturalness and fluency. The audio samples and code are available at \url{https://github.com/Ai-S2-Lab/FluentEditor}.
\end{abstract}
%

\vspace{-2mm}
\section{Introduction}
\label{sec:intro} 
\vspace{-2mm}
Text-based speech editing (TSE) \cite{jin2017voco} allows for modification of the output audio by editing the transcript rather than the audio itself. With the rapid development of the internet, audio-related media sharing has become a prevalent activity in our daily lives. Note that TSE can bring great convenience to the audio generation process and be applied to a variety of areas with personalized voice needs, including video creation for social media, games, and movie dubbing.

Over the past few years, many attempts adopted text-to-speech (TTS) systems to build neural network-based TSE models. For example, the CampNet \cite{wang2022campnet} conducts mask training on a context-aware neural network based on Transformer to improve the quality of the edited voice. $A^3T$ \cite{bai20223} suggests an alignment-aware acoustic and text pretraining method, which can be directly applied to speech editing by reconstructing masked acoustic signals through text input and acoustic text alignment.
More recently, the diffusion model has gradually become the backbone of the NN-based TSE with remarkable results. For example, 
EdiTTS \cite{tae2021editts} takes the diffusion-based TTS model as the backbone and proposes a score-based TSE methodology for fine-grained pitch and content editing.
FluentSpeech \cite{jiang2023fluentspeech} proposes a context-aware diffusion model that iteratively refines the modified mel-spectrogram with the guidance of context features.

However, during training, the existing approaches just constrain the Euclidean Distance \cite{reyes2016spectrum} between the mel-spectrum to be predicted and the ground truth to ensure the naturalness of TSE. Although they consider the use of contextual information to mitigate the over-smoothing problem of edited speech, their objective functions are not designed to ensure fluent output speech~\cite{tseng2005fluent, liu2021expressive}.
We consider two challenges to be tackled for effective speech fluency modeling. 1) \textit{Acoustic Consistency}: the smoothness of the concatenation between the region to be edited and its neighboring regions should be close to the real concatenation point \cite{hunt1996unit}.
2) \textit{Prosody Consistency}: the prosody style of the synthesized audio in the region to be edited needs to be consistent with the prosody style of the original utterance \cite{raitio2020controllable, wang2018style}.

To address the above issues, we propose a novel fluency speech editing scheme, termed FluentEditor, by introducing the acoustic and prosody consistency training criterion to achieve natural and fluent speech editing. Specifically, 1) To achieve the acoustic consistency, we design the \textit{Acoustic Consistency Loss} $\mathcal{L}_{AC}$ to calculate whether the variance at the boundaries is close to the variance at the real concatenation points. 2) To achieve the prosody consistency, we introduce the \textit{Prosody Consistency Loss} $\mathcal{L}_{PC}$ to let the high-level prosody features of the synthesized audio in the region to be edited be close to that of the original utterance. The high-level prosody features are extracted by the pre-trained GST-based prosody extractor\cite{wang2018style}. The subjective and objective results on the VCTK \cite{veaux2017cstr} dataset show that the acoustic and prosody consistency of the FluentEditor is significantly better than the advanced TSE baselines, while the proposed FluentEditor can ensure a high degree of fluency like real speech. 

The main contributions of this work can be summarized as follows:
1) We propose a novel fluency speech editing scheme, termed FluentEditor; 2) We adopt the diffusion model as the backbone and introduce \textit{Acoustic and Prosody Consistency Losses} to conduct the fluency modeling for TSE; 3) The proposed model outperforms all advanced TSE baselines in terms of naturalness and fluency.




 

 \vspace{-2mm}
\section{FluentEditor: Methodology}
\label{sec:model}
\vspace{-2mm}

We formulate the proposed FluentEditor, a TSE model that ensures speech fluency by considering acoustic and prosody consistency. We first introduce the overall workflow, then further elaborate the fluency-aware training criterion and the run-time inference.

 \vspace{-3mm}
\subsection{Overall Workflow}
\begin{figure*}[!t]
\centering
\setlength{\abovecaptionskip}{-0mm}   
\centerline{\includegraphics[width=0.98\linewidth]{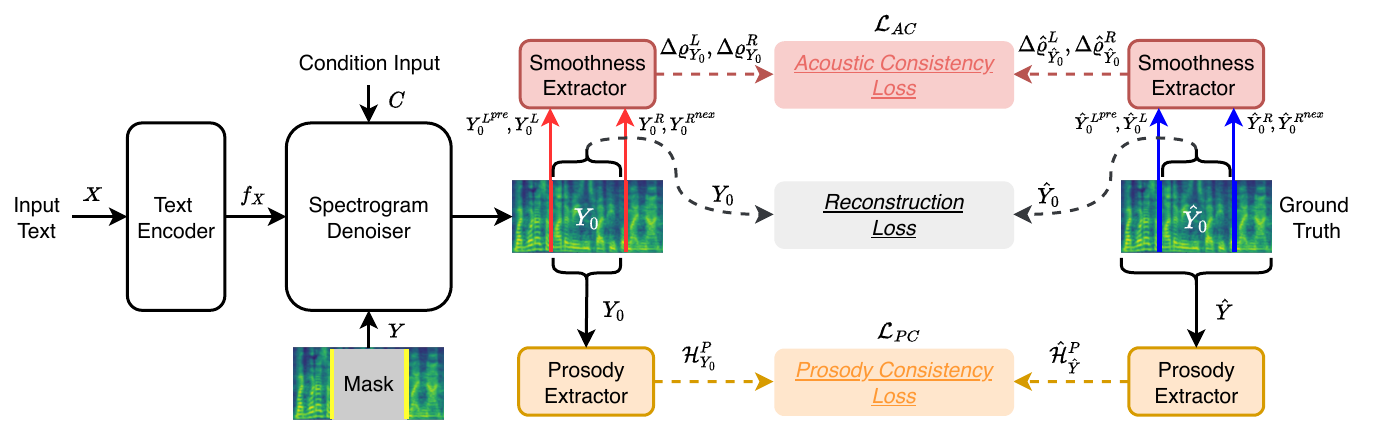}}
\vspace{-3mm}
\caption{The overall workflow of FluentEditor. The total loss function includes Reconstruction Loss, and Acoustic and Prosody Consistency Losses.}
\vspace{-4mm}
\label{fig:fig1}
\end{figure*}

\vspace{-2mm}
As shown in Fig.\ref{fig:fig1}, our FluentEditor adopts the mask prediction-based diffusion network as the backbone, which consists of a text encoder, and a spectrogram denoiser. The spectrogram denoiser seeks to adopt the Denoising diffusion probabilistic model (DDPM) to learn a data distribution $p(\cdot)$ by gradually denoising a normally distributed variable through the reverse process of a fixed Markov Chain of length $T$.

Assume that the phoneme embedding of the input phoneme sequence is $X = (X_1, \ldots , X_{|X|})$ and the acoustic feature sequence for $X$ is $\hat Y = (\hat Y_1, \ldots , \hat Y_{|\hat Y|})$. The masked acoustic feature sequence $\hat{Y}_{mask} = Mask(\hat Y, \lambda)$ is obtained by replacing the random spans of $\hat Y$ with the random vector according to a $\lambda$ probability.
Specifically, the text encoder aims to extract the high-level linguistic feature $\mathcal{H}_{X}$ for $X$. The spectrogram denoiser then aggregates the $\mathcal{H}_{X}$ and the condition input $C$ to guide the reverse process of the diffusion model $\Theta(Y_{t}|t, C)$ ($t \in T$), where $Y_{t}$ is a noisy version of the clean input $\hat{Y}_{0}$. Similar to \cite{jiang2023fluentspeech}, the condition input $C$ consists of the frame-level linguistic feature $\mathcal{H}^{f}_{X}$, acoustic feature sequence $\hat Y$, masked acoustic feature sequence $\hat{Y}_{mask}$, speaker embedding $e_{spk}$ and the pitch embedding $e_{pitch}$. 
In the generator-based diffusion models, $p_{\theta}(Y_{0}|Y_{t})$ is the implicit distribution imposed by the neural network $f_{\theta}(Y_t, t)$ that outputs $Y_{0}$ given $Y_{t}$. And then $Y_{t-1}$ is sampled using the posterior distribution $q(Y_{t-1} | Y_{t}, Y_{0})$ given $Y_{t}$ and the predicted $Y_{0}$.


To model speech fluency, we design \textit{acoustic consistency loss} $\mathcal{L}_{AC}$ and \textit{prosody consistency loss} $\mathcal{L}_{PC}$ on the basis of the original \textit{reconstruction loss}, to ensure that the acoustic and prosody performance of speech generated in the editing area is consistent with the context and the original utterance. For reconstruction loss,  we follow \cite{jiang2023fluentspeech} and employ Mean Absolute Error  (MAE) and the Structural Similarity Index (SSIM) \cite{ren2022revisiting} losses to calculate the difference between $Y_{0}$ and the corresponding ground truth segment $\hat Y_{0}$. 
In the following subsection, we will introduce $\mathcal{L}_{AC}$ and $\mathcal{L}_{PC}$ in detail.


\vspace{-3mm}
\subsection{Fluency-Aware Training criterion}
\vspace{-2mm}


\subsubsection{Acoustic Consistency Loss}
\vspace{-2mm}

The acoustic consistency loss $\mathcal{L}_{AC}$ employs smoothness constraints at both the left and right boundaries for the predicted acoustic feature $Y_{0}$. We compare the variance, $\Delta \varrho_{Y_0}^L$ and $\Delta \varrho_{Y_0}^R$, for the left and right boundaries with  $\Delta \hat{\varrho}_{\hat{Y_0}}^{L/R}$ of ground truth speech at the corresponding boundaries to serve as the proxy of the overall smoothness $\mathcal{L}_{AC}$. 

Specifically, $\mathcal{L}_{AC}$ consists of $\mathcal{L}_{AC}^L$ and $\mathcal{L}_{AC}^R$, and we use Mean Squared Error (MSE)\cite{mandhare2019generalized} to measure the proximity between the target segment and the ground truth:
\begin{equation}
\begin{aligned}
\mathcal{L}_{AC} &=\mathcal{L}_{AC}^L +\mathcal{L}_{AC}^R \\&= \rm{MSE}(\Delta \varrho_{Y_0}^L, \Delta \hat{\varrho}_{\hat{Y_0}}^L) + \rm{MSE}(\Delta \varrho_{Y_0}^R, \Delta \hat{\varrho}_{\hat{Y_0}}^R)
\end{aligned} 
 \vspace{-2mm}
\end{equation}


Note that the Euclidean distance between two adjacent frames is obtained by the smoothness extractor. Take $\Delta \varrho_{Y_0}^L$ as an example, 
\begin{equation} 
\vspace{-1mm}
\Delta \varrho_{Y_0}^L = \varrho_{{Y_0}^L} -\varrho_{Y^{{L}^{pre}}_{0}}
 \vspace{-2mm}
\end{equation}
where $Y^{{L}^{pre}}_{0}$ denotes the speech frame preceding the left boundary of the masked region. In other words, the ending frame of the adjacent non-masked region is on the left side.
To comprehensively capture the statistical properties of the audio signal, we utilize variance to describe the feature information of each Mel spectrogram frame, denoted as $\varrho_{{Y_0}^L}$ and $\varrho_{Y^{{L}^{pre}}_{0}}$.

Similarly, we compute the smoothness constraint for the right boundary$\mathcal{L}_{AC}^R$, where $Y^{{R}^{nex}}_{0}$ denotes the speech frame succeeding the right boundary of the masked region, in other words, the starting frame of the adjacent non-masked region on the right side.

\vspace{-2mm}
\subsubsection{Prosody Consistency Loss}
\vspace{-2mm}
The prosody consistency loss $\mathcal{L}_{PC}$ is responsible for capturing the prosody feature $\mathcal{H}_{Y_0}^P$ from the predicted region $Y_0$ while also analyzing the overall prosody characteristics $\hat{\mathcal{H}}_{\hat{Y}}^P$ present in the original speech, then employ the MSE loss to conduct the prosody consistency constraints.
\begin{equation} 
\vspace{-1mm}
\mathcal{L}_{PC} = \text{MSE}(\mathcal{H}_{Y_0}^P, \hat{\mathcal{H}}_{\hat{Y}}^P) 
\vspace{-2mm}
\end{equation}

Note that the prosody features $\mathcal{H}_{Y_0}^P$ and $\hat{\mathcal{H}}_{\hat{Y}}^P$ are obtained by the pre-trained prosody extractor. 
Specifically, the prosody extractor utilizes the reference encoder \cite{wang2018style} of the Global Style Token (GST) \cite{wang2018style} model to convert $Y_0$ and $\hat{Y}$ into high-level prosody features with fixed length for easy comparison.
\begin{equation} 
\mathcal{H}_{Y_0}^P = \text{GST}(Y_0), \ \ \ \hat{\mathcal{H}}_{\hat{Y}}^P = \text{GST}(\hat{Y})
\vspace{-1mm}
\end{equation}


Lastly, following \cite{jiang2023fluentspeech}, the total loss function is the sum of reconstruction loss and two new loss functions, $\mathcal{L}_{AC}$ and $\mathcal{L}_{PC}$, across all non-contiguous masked regions, since the mask region in a sentence may include multiple non-contiguous segments \cite{jiang2023fluentspeech}. In a nutshell, $\mathcal{L}_{AC}$ and $\mathcal{L}_{PC}$ of the FluentEditor are introduced to ensure fluent speech with consistent prosody.

\vspace{-3mm}
\subsection{Run-time Inference}
\vspace{-2mm}
In run-time, given the original text and its speech, the user can edit the speech by editing the text. Note that we can manually define modification operations (i.e., insertion, replacement, and deletion). The corresponding speech segment of the edited word in the given text is treated as the masked regions in Fig. \ref{fig:fig1}. Similar to \cite{jiang2023fluentspeech}, our FluentEditor reads the edited text and the remaining acoustic feature $\hat{Y} - \hat{Y}_{mask}$ of the original speech to predict the $Y_{0}$ for the edited word. At last, the $Y_{0}$ and its context $\hat{Y} - \hat{Y}_{mask}$ are concatenated as the final fluent output speech.

\begin{table*}[t]
    \centering
    \setstretch{0.88}
    \caption{Objective and subjective evaluation results of comparative study. * means the value achieves suboptimal.}
    \begin{tabular}{p{3.7cm}<{\centering}|p{2cm}<{\centering}p{2cm}<{\centering}p{2cm}<{\centering}|p{2.5cm}<{\centering}p{2.5cm}<{\centering}}
        \toprule
        \multirow{2}{*}{\textbf{Method}} & \multicolumn{3}{c|}{\textbf{Objective Evaluation}} &  \multicolumn{2}{c}{\textbf{Subjective Evaluation (FMOS)}} \\
         & \textbf{MCD} $(\downarrow)$   & \textbf{STOI} $(\uparrow)$  & \textbf{PESQ} $(\uparrow)$  &  \multicolumn{1}{c}{\textbf{Insertion}}  & \multicolumn{1}{c}{\textbf{Replacement}}  \\
        \midrule
        Ground Truth & NA & NA & NA & 4.37 $\pm$ 0.05 & 4.42 $\pm$ 0.01 \\ \hline
       
        CampNet & 3.85 & 0.53 & 1.38 & 3.89 $\pm$ 0.01 & 3.94 $\pm$ 0.03\\
        $A^3T$ & 3.79 & 0.76 & 1.59 & 3.82 $\pm$ 0.03 & 3.83 $\pm$ 0.02\\  
        FluentSpeech & 3.50 & 0.79 &  1.93 & 4.02 $\pm$ 0.04 & 4.04 $\pm$ 0.01\\ \hline
        \textbf{FluentEditor (Ours)}  & \textbf{3.47} & \textbf{0.81} & \textbf{1.85*} & \textbf{4.25 $\pm$ 0.03} & \textbf{4.26 $\pm$ 0.01} \\

        \bottomrule
    \end{tabular}
    \vspace{-6mm}
    \label{tab1}
\end{table*}

\begin{table}
    \centering
    \caption{Objective and subjective results of ablation study.}
    \begin{tabular}{p{3.1cm}<{\centering}|p{1.8cm}<{\centering}p{1.8cm}<{\centering}}
        \toprule
        \textbf{Method} & \textbf{C-FMOS}  & \textbf{MCD} $(\downarrow)$  \\
        \midrule
        FluentEditor & \textbf{0.00} & \textbf{3.47} \\
       \quad w/o $ \mathcal{L}_{AC}$ & -0.16 & 3.48 \\
          \quad  w/o $ \mathcal{L}_{PC}$ & -0.21 & 3.51 \\
        \bottomrule
    \end{tabular}
    \vspace{-4mm}
    \label{ablation study}
\end{table}

\vspace{-4mm}
\section{Experiments and Results}
\label{sec:exp}
\vspace{-2mm}
\subsection{Dataset}
\vspace{-2mm}
We validate the FluentEditor on the VCTK \cite{veaux2017cstr} dataset, which is an English speech corpus uttered by 110 English speakers with various accents. Each recording is sampled at 22050 Hz with 16-bit quantization. The precise forced alignment is achieved through Montreal Forced Aligner (MFA) \cite{mcauliffe2017montreal}. We partition the dataset into training, validation, and testing sets, randomly with 98\%, 1\%, and 1\%, respectively.
 \vspace{-2mm}
\subsection{Experimental Setup}
\vspace{-2mm}
The configurations of text encoder and spectrogram denoiser are referred to \cite{jiang2023fluentspeech}.
The diffusion steps $T$ of the FluentEditor system is set to 8.
Following GST \cite{wang2018style}, the prosody extractor comprises a convolutional stack and an RNN. The dimension of the output prosody feature of the GST-based prosody extractor is 256.
Following \cite{bai20223}, we adopt a random selection strategy, with a fixed masking rate of 80\%, for masking specific phoneme spans along with their corresponding speech frames.
The pre-trained HiFiGAN \cite{kong2020hifi} vocoder is used to synthesize the speech waveform.
We set the batch size is 16. The initial learning rate is set at $2 \times 10^{-4}$, and the Adam optimizer \cite{kingma2014adam} is utilized to optimize the network.
The FluentEditor model is trained with 2 million training steps on one A100 GPU. 
\vspace{-2mm}
\subsection{Evaluation Metric}
\vspace{-2mm}
For subjective evaluation, We conduct a Mean Opinion Score (MOS) \cite{loizou2011speech} listening evaluation in terms of speech fluency, termed \textit{FMOS}. 
Note that FMOS allows the listener to feel whether the edited segments of the edited speech are fluent compared to the context. We keep the text content and text modifications consistent among different models to exclude other interference factors, only examining speech fluency.
Furthermore, Comparative FMOS (C-FMOS) \cite{loizou2011speech} is also used to conduct the ablation study. 
For objective evaluation, we utilize MCD \cite{kubichek1993mel}, STOI \cite{taal2010short}, and PESQ \cite{rix2001perceptual} to measure the overall quality of the edited speech.
 \vspace{-2mm}
\subsection{Comparative Study}
\vspace{-2mm}
We develop four neural TSE systems for a comparative study, that includes: 1) \textbf{CampNet} \cite{wang2022campnet} propose a context-aware mask prediction network to simulate the process of text-based speech editing; 2) $\boldsymbol{A^3T}$ \cite{bai20223} propose the alignment-aware acoustic-text pre-training that takes both phonemes and partially-masked spectrograms as inputs; 3) \textbf{FluentSpeech} \cite{jiang2023fluentspeech} takes the diffusion model as backbone and predict the masked feature with the help of context speech; and 4) \textbf{FluentEditor (Ours)} designs the acoustic and prosody consistency losses. We also add the \textbf{Ground Truth} speech for comparison. Note that two ablation systems, that are ``$\boldsymbol{w/o}$ $\boldsymbol{\mathcal{L}_{AC}}$'' and ``$\boldsymbol{w/o}$ $\boldsymbol{\mathcal{L}_{PC}}$'', are built to validate the two new losses.

\vspace{-2mm}
\subsection{Main Results}
\vspace{-2mm}
\textbf{Objective results:} We select 400 test samples from the test set randomly and report the objective results in the second to fourth columns of Table \ref{tab1}. Note that we follow \cite{jiang2023fluentspeech} and just measure the objective metrics of the masked region using the reconstructed speech. We observe that our FluentEditor achieves the best performance in terms of overall speech quality. For example, the MCD and STOI values of FluentEditor obtain optimal results and PESQ achieves suboptimal results among all systems. It suggests that the FluentEditor performs proper acoustic feature prediction for the speech region to be edited. 
Note that objective metrics do not fully reflect the human perception \cite{ren2020fastspeech}, we further conduct subjective listening experiments.

\noindent \textbf{Subjective results:} For FMOS evaluation, we selected 50 audio samples from the test set and invited 20 listeners to evaluate speech fluency. Following \cite{tan2021editspeech}, we test the insertion and replacement operations and present the FMOS results in the last two columns of Table \ref{tab1}. 
We find that FluentEditor consistently achieves superior fluency-related perceptual scores. For example, FluentEditor obtains the top FMOS value of 4.25 for insertion and 4.26 for replacement, that very close to that of ground truth.
This demonstrates the effectiveness of the fluency-aware training criterion. By considering the acoustic and prosody consistency constraints, our FluentEditor allows for weakening the editing traces and improving the prosody performance of the edited speech.

\vspace{-4mm}
\subsection{Ablation Study}
\vspace{-2mm}
To further validate the contribution of our $\mathcal{L}_{AC}$ and $\mathcal{L}_{PC}$ respectively, the subjective and objective ablation results, of insertion and replacement, are reported in Table \ref{ablation study}. We follow the previous section to prepare the samples and listeners. It's observed that the C-FMOS and MCD values of these two ablation systems both drop when we remove the $\mathcal{L}_{AC}$ and $\mathcal{L}_{PC}$ respectively, indicating that the acoustic and prosody consistency constraints play a vital role in enhancing both the naturalness and fluency of the edited speech.

\begin{figure}[!t]
\centering
\setlength{\abovecaptionskip}{-0mm}   
\begin{minipage}{0.7\linewidth}
  \centerline{
  \includegraphics[width= \linewidth]{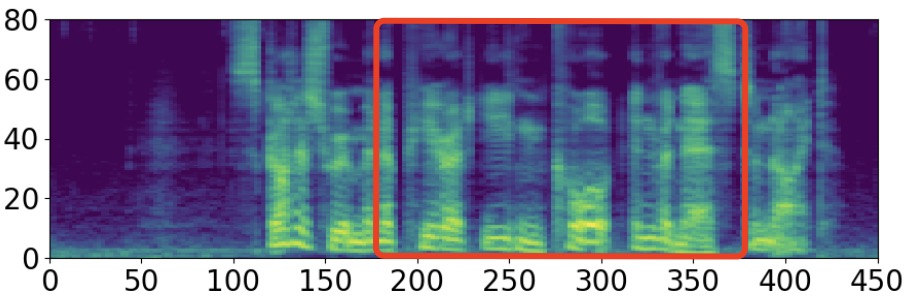}
  }
  \vspace{-0.5mm}
  \centerline{(a) {\small FluentSpeech} }
\end{minipage}
\vfill
\begin{minipage}{0.7\linewidth}
  \centerline{
  \includegraphics[width= \linewidth]{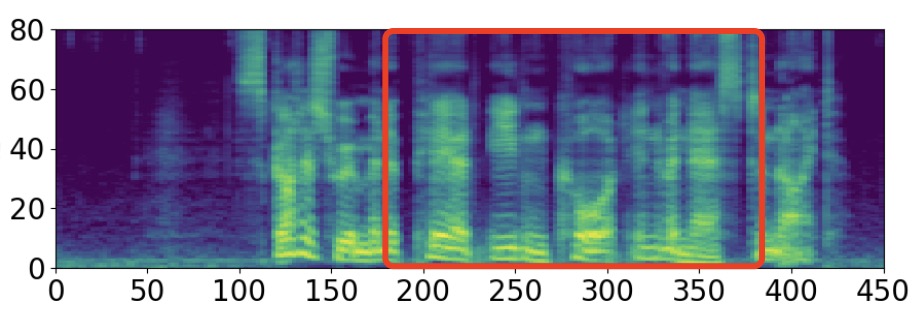}
  }
  \vspace{-0.5mm}
  \centerline{(b) {\small FluentEditor} }
\end{minipage}
\vspace{-2.6mm}
\caption{Visualizations of the generated mel-spectrograms by FluentEditor and FluentSpeech baseline. 
}
\label{fig:spec}
\vspace{-5mm}
\end{figure}

 \vspace{-3mm}
\subsection{Visualization Analysis}
\vspace{-2mm}
As illustrated in the Fig.\ref{fig:spec}, we visualize the mel-spectrograms produced by FluentEditor and the FluentSpeech baseline\footnote{Due to space limits, we just report the FluentSpeech baseline. More visualization results and speech samples are referred to our website: \url{https://github.com/Ai-S2-Lab/FluentEditor}. }.
The red boxes indicate the random masked and its reconstructed speech segment of utterance ``Scottish Women appear at Eden Court, Inverness, tonight.''. 
We can see that FluentEditor can generate mel-spectrograms with richer frequency details compared with the baseline, resulting in natural and expressive sounds, which further demonstrates the effectiveness of acoustic and prosody consistency losses. Nevertheless, we recommend that the reader listen to our speech samples$^1$ to visualize the advantages.

  \vspace{-4mm}
\section{Conclusion}
\label{sec:con}
 \vspace{-4mm}
In this paper, we introduce a novel text-based speech editing (TSE) model, termed FluentEditor, that involves two novel fluency-aware training criterions to improve the acoustic and prosody consistency of edited speech. 
The acoustic consistency loss $\mathcal{L}_{AC}$ to calculate whether the variance at the boundaries is close to the variance at the real concatenation points, while the prosody consistency loss $\mathcal{L}_{PC}$ to let the high-level prosody features of the synthesized audio in the region to be edited be close to that of the original utterance.
The objective and subjective experiments on VCTK demonstrate that incorporating $\mathcal{L}_{AC}$ and $\mathcal{L}_{PC}$ yields superior results and ensures fluent speech with consistent prosody. In future work, we will consider the multi-scale consistency and further improve the FluentEditor architecture.

 

 
\bibliographystyle{IEEEbib}
{\footnotesize
\bibliography{strings}}

\end{document}